\documentclass[prd,nofootinbib,twocolumn,showpacs,preprintnumbers,floatfix,amsmath,amssymb,amsfonts]{revtex4}

\usepackage{graphicx}
\usepackage{amsmath}
\usepackage{dcolumn}
\usepackage{epsfig}
\usepackage{psfrag}
\usepackage{lineno}
\usepackage{color}

\RequirePackage{xspace}

\DeclareGraphicsExtensions{.epsi,.eps,.ps,.eps.gz,.ps.gz,.gif,.epsi.gz}
\RequirePackage{xspace}

\usepackage{relsize}
\def\babar{\mbox{\slshape B\kern-0.1em{\smaller A}\kern-0.1em
    B\kern-0.1em{\smaller A\kern-0.2em R}}\xspace}

\def\belle{Belle\xspace}

\def\epem       {\ensuremath{e^+e^-}\xspace}

\def\cbar  {\ensuremath{\overline c}\xspace}

\def\piz   {\ensuremath{\pi^0}\xspace}

\def\kaon  {\ensuremath{K}\xspace}
\def\Kbar  {\kern 0.2em\overline{\kern -0.2em K}{}\xspace}
\def\Kb    {\ensuremath{\Kbar}\xspace}
\def\KKb   {\ensuremath{\kaon {\kern -0.16em \Kb}}\xspace}
\def\Kz    {\ensuremath{K^0}\xspace}
\def\Kzb   {\ensuremath{\Kbar^0}\xspace}
\def\KzKzb {\ensuremath{\Kz \kern -0.16em \Kzb}\xspace}
\def\Kp    {\ensuremath{K^+}\xspace}
\def\Km    {\ensuremath{K^-}\xspace}

\def\KpKm  {\ensuremath{\Kp \kern -0.16em \Km}\xspace}
\def\KS    {\ensuremath{K^0_{\scriptscriptstyle S}}\xspace} 
\def\KSL   {\ensuremath{K^0_{\scriptscriptstyle S, L}}\xspace} 
\def\KL    {\ensuremath{K^0_{\scriptscriptstyle L}}\xspace}

\def\rhoz    {\ensuremath{\rho^0}\xspace}

\def\Mz      {\ensuremath{M^0}\xspace}
\def\Mbar    {\kern 0.2em\overline{\kern -0.2em M}{}\xspace}
\def\Mzb     {\ensuremath{\Mbar^0}\xspace}

\def\D       {\ensuremath{D}\xspace}
\def\Dbar    {\kern 0.2em\overline{\kern -0.2em D}{}\xspace}
\def\Db      {\ensuremath{\Dbar}\xspace}
\def\Dz      {\ensuremath{D^0}\xspace}
\def\Dzb     {\ensuremath{\Dbar^0}\xspace}
\def\DzDzb   {\ensuremath{\Dz {\kern -0.16em \Dzb}}\xspace}
\def\Dp      {\ensuremath{D^+}\xspace}
\def\Dm      {\ensuremath{D^-}\xspace}

\def\DpDm    {\ensuremath{\Dp {\kern -0.16em \Dm}}\xspace}
\def\Dstar   {\ensuremath{D^*}\xspace}
\def\Dstarb  {\ensuremath{\Dbar^*}\xspace}

\def\DDb     {\ensuremath{\D {\kern -0.16em \Db}}\xspace}

\def\DstarDstarb     {\ensuremath{\Dstar{\kern -0.16em \Dstarb}}\xspace}

\def\B       {\ensuremath{B}\xspace}
\def\Bbar    {\kern 0.18em\overline{\kern -0.18em B}{}\xspace}

\def\Bz      {\ensuremath{B^0}\xspace}
\def\Bzb     {\ensuremath{\Bbar^0}\xspace}
\def\BzBzb   {\ensuremath{\Bz {\kern -0.16em \Bzb}}\xspace}
\def\BzBz    {\ensuremath{\Bz {\kern -0.16em \Bz}}\xspace}
\def\BzbBzb  {\ensuremath{\Bzb {\kern -0.16em \Bzb}}\xspace}
\def\Bu      {\ensuremath{B^+}\xspace}
\def\Bub     {\ensuremath{B^-}\xspace}

\def\BpBm    {\ensuremath{\Bu {\kern -0.16em \Bub}}\xspace}
\def\Bs      {\ensuremath{B^0_s}\xspace}
\def\Bsb     {\ensuremath{\Bbar^0_s}\xspace}
\def\BsBsb   {\ensuremath{\Bs {\kern -0.08em \Bsb}}\xspace}

\def\P       {\ensuremath{P}\xspace}
\def\Pbar    {\kern 0.18em\overline{\kern -0.18em P}{}\xspace}

\def\BorBbar    {\kern 0.18em\optbar{\kern -0.18em B}{}\xspace}
\def\DorDbar    {\kern 0.18em\optbar{\kern -0.18em D}{}\xspace}
\def\KorKbar    {\kern 0.18em\optbar{\kern -0.18em K}{}\xspace}

\mathchardef\Upsilon="7107
\def\Y#1S{\ensuremath{\Upsilon{(#1S)}}\xspace}
\def\FourS{\ensuremath{\Upsilon{(4S)}}\xspace}
\def\FiveS{\ensuremath{\Upsilon{(5S)}}\xspace}

\def\chib#1  {\ensuremath{\chi_{b#1}\xspace}}

\newcommand{\spect}[4][]{\ensuremath{\ifthenelse{\equal{#1}{}} {} {#1\,} {}^{#2\!} {#3}_{#4}}}

\mathchardef\Deltares="7101
\mathchardef\Xi="7104
\mathchardef\Lambda="7103
\mathchardef\Sigma="7106
\mathchardef\Omega="710A

\def\Deltabar {\kern 0.25em\overline{\kern -0.25em \Deltares}{}\xspace}
\def\Lbar     {\kern 0.2em\overline{\kern -0.2em\Lambda\kern 0.05em}\kern-0.05em{}\xspace}
\def\Sigbar   {\kern 0.2em\overline{\kern -0.2em \Sigma}{}\xspace}
\def\Xibar    {\kern 0.2em\overline{\kern -0.2em \Xi}{}\xspace}
\def\Obar     {\kern 0.2em\overline{\kern -0.2em \Omega}{}\xspace}
\def\Nbar     {\kern 0.2em\overline{\kern -0.2em N}{}\xspace}
\def\Xb       {\kern 0.2em\overline{\kern -0.2em X}{}\xspace}

\newcommand{\zev}{\ensuremath{\mathrm{\,Ze\kern -0.1em V}}\xspace}
\newcommand{\eev}{\ensuremath{\mathrm{\,Ee\kern -0.1em V}}\xspace}
\newcommand{\pev}{\ensuremath{\mathrm{\,Pe\kern -0.1em V}}\xspace}
\newcommand{\tev}{\ensuremath{\mathrm{\,Te\kern -0.1em V}}\xspace}
\newcommand{\gev}{\ensuremath{\mathrm{\,Ge\kern -0.1em V}}\xspace}
\newcommand{\mev}{\ensuremath{\mathrm{\,Me\kern -0.1em V}}\xspace}
\newcommand{\kev}{\ensuremath{\mathrm{\,ke\kern -0.1em V}}\xspace}
\newcommand{\ev}{\ensuremath{\mathrm{\,e\kern -0.1em V}}\xspace}
\newcommand{\gevc}{\ensuremath{{\mathrm{\,Ge\kern -0.1em V\!/}c}}\xspace}
\newcommand{\mevc}{\ensuremath{{\mathrm{\,Me\kern -0.1em V\!/}c}}\xspace}
\newcommand{\gevcc}{\ensuremath{{\mathrm{\,Ge\kern -0.1em V\!/}c^2}}\xspace}
\newcommand{\mevcc}{\ensuremath{{\mathrm{\,Me\kern -0.1em V\!/}c^2}}\xspace}

\def\invab   {\ensuremath{\mbox{\,ab}^{-1}}\xspace}

\definecolor{Red}{rgb}{1,0,0}
\definecolor{Green}{rgb}{0,1,0}
\definecolor{Blue}{rgb}{0,0,1}
\definecolor{Black}{rgb}{0,0,0}

\def\to                 {\ensuremath{\rightarrow}\xspace}

\def\pep2{PEP-II}

\def\BFs{$B$ Factories\xspace}

\def\gsim{{~\raise.15em\hbox{$>$}\kern-.85em
          \lower.35em\hbox{$\sim$}~}\xspace}
\def\lsim{{~\raise.15em\hbox{$<$}\kern-.85em
          \lower.35em\hbox{$\sim$}~}\xspace}

\def\CP      {\ensuremath{C\!P}\xspace}

\def\CPT     {\ensuremath{C\!PT}\xspace}
\def\C       {\ensuremath{C}\xspace}
\def\P       {\ensuremath{P}\xspace}
\def\T       {\ensuremath{T}\xspace}

 \def\p1b{\ensuremath{\phi_1}\xspace}

\def\deltat   {\ensuremath{{\rm \Delta}t}\xspace}
\def\deltam   {\ensuremath{{\rm \Delta}m}\xspace}

\xspace

\def\jetset74   {\mbox{\tt Jetset \hspace{-0.5em}7.\hspace{-0.2em}4}\xspace}

\long\def\inst#1{\par\nobreak\kern 4pt\nobreak
    {\it #1}\par\vskip 10pt plus 3pt minus 3pt}

\begin{document}


{\pagestyle{empty}

\par\vskip 3cm

\title{
\Large \boldmath
  Neutral meson tests of time-reversal symmetry invariance
  }

\author{A. J. Bevan}
\author{G. Inguglia}
\author{M. Zoccali}
\affiliation{Queen Mary University of London, Mile End Road, E1 4NS, United Kingdom}

\date{\today}

\begin{abstract}
%
%
The laws of quantum physics can be studied under the mathematical operation \T
that inverts the direction of time. Strong and electromagnetic forces are 
known to be invariant under temporal inversion, however the weak force is 
not. The \babar experiment recently exploited the quantum-correlated production 
of pairs of \Bz mesons to show that \T is a broken symmetry. Here we show that it is 
possible to perform a wide range of tests under 
\T (as well as \CP and \CPT) in order to validate the Standard Model of particle physics 
covering \CP filter states with $b$ to $u$, $d$, $s$, and $c$ transitions as well as $c$ to $d$ and $s$
transitions using entangled \B and \D pairs created in \FourS and $\psi(3770)$
decays.  We also note that pseudoscalar decays to two spin one particle final states
provide an additional set of \CP filter bases to use for such tests.

\end{abstract}


\maketitle

\vfill

}

\setcounter{footnote}{0}

Weak decays are known to violate the set of discrete symmetries charge conjugation (\C), 
spatial inversion otherwise known as parity (\P)~\cite{Wu:1957my}, \CP~\cite{Christenson:1964fg} which distinguishes between matter and 
antimatter, and time-reversal (\T)~\cite{CPLEAR,Lees:2012kn}.  
Historically there have been a number of incorrect claims of testing \T violation through 
triple product asymmetries of four body decays of $K$, $B$, and $D$ decays.
Such approaches are not able to test \T symmetry invariance as one is not able 
to experimentally identify an asymmetry of \T conjugate transitions such as 
that indicated in Eq.~(\ref{EQ:TASYM}). One can find a recent review of triple
product asymmetry measurements using the correct nomenclature in Ref.~\cite{Gronau:2011cf}.
In this paper we identify a new set of orthogonal \CP filter bases that can 
be used for \T violation tests.  Using these, along with \KS/\KL basis filters identified
in  Ref.~\cite{Banuls:1999aj,Banuls:2000ki,Alvarez:2006nk,Bernabeu:2012ab},
we outline a programme of symmetry invariance tests in \B and \D decays. 
We also illustrate how these measurements relate to the SM weak interaction quark mixing 
mechanism given by the Cabibbo-Kobayashi-Maskawa (CKM) matrix in the SM~\cite{Cabibbo:1963yz,Kobayashi:1973fv},
where the Kobayashi-Maskawa phase $\delta_{KM}$ is responsible for both \CP
and \T violation in the SM.

Experimental evidence so far supports the hypothesis that the overall 
symmetry \CPT is conserved, for example see~\cite{Aubert:2003hd,Aubert:2004xga,Aubert:2006nf,Aubert:2007bp,Higuchi:2012kx} for the results of tests using \B meson decays.  The most significant hint for a non-conservation of \CPT
comes from \babar, however the \CPT violation search using sidereal time evolution of di-lepton
decays only has a significance of $2.8\sigma$ from expectations of being consistent with the
Standard Model of particle physics (SM)~\cite{Aubert:2007bp}.
For \CPT to be conserved, the level of \CP violation has to be balanced 
by `just enough' \T violation so that these two effects cancel each other out to preserve 
\CPT symmetry.
 While this discussion may seem to be of academic interest it is worth recalling 
that \CPT is conserved in locally invariant field theories such as the 
SM~\cite{Luders:1954zz}.  A number of scenarios that have been 
proposed in order to work toward a theory of quantum gravity can naturally 
manifest \CPT violation.  The details of how this might happen depend on the specifics
of such models (see for example~\cite{Kostelecky:2002pf,Bernabeu:2003ym,Bernabeu:2006av}), and as the 
weak force has a lack of respect for discrete symmetries one should be motivated to 
test the behaviour of this force with respect to \T (and other symmetries) in as many 
types of weak decay as possible to experimentally verify if the results are found to be 
consistent with expectations.  Such a programme of measurements would parallel the time-dependent \CP
violation measurements performed by \babar, Belle, the Tevatron, and LHC experiments
since the start of this millennium.

The observable used to study \T symmetry invariance parallels that of a time-dependent 
or time-integrated \CP asymmetry; one constructs a rate asymmetry of \T conjugate processes
from some state $|1\rangle$ to another $|2\rangle$ state.  
Under \T the time ordering of these two states is reversed, i.e.
\begin{equation}
A_\T = \frac{\Gamma(|1\rangle \to |2\rangle ) - \Gamma(|2\rangle \to |1\rangle ) } {\Gamma(|1\rangle \to |2\rangle ) + \Gamma(|2\rangle \to |1\rangle )}, \label{EQ:TASYM}
\end{equation}
and \T symmetry invariance is violated if $A_T \neq 0$. As strong and electromagnetic processes
conserve \T, one needs to identify weak transitions that satisfy Eq.~(\ref{EQ:TASYM}).
It is not trivial to identify measurable systems that satisfy this condition, and the 
real issue that remains is to identify the sets of interesting states,
that can be used for such a test.  The transition $|1\rangle \to |2\rangle$
requires a \T conjugate partner $|2\rangle \to |1\rangle$ that 
is experimentally distinguishable.

A key to identifying pairs of measurable \T conjugate states is highlighted 
in Ref.~\cite{Bernabeu:2012ab}.  One studies the ensemble of entangled
[or Einstein-Podolsky-Rosen (EPR) correlated]~\cite{EPR,Bell:1964kc} neutral meson pairs, 
and reconstructs an experimentally distinguishable double tagged \T conjugate pair 
of transitions.  If one follows this procedure using filter decays into only one orthonormal
basis it is possible to construct an asymmetry that is both \T and \CP violating.  For 
example this parallels the concept of a Kabir asymmetry measured in kaon decays~\cite{Kabir:1970ts}. 
The \BFs have performed such tests of the dual asymmetries $A_{CP,T}$ and $A_{CP,CPT}$ using
dilepton and hadronic final states~\cite{Aubert:2002mn,Aubert:2003hd,Aubert:2004xga,Nakano:2005jb,Aubert:2006nf}, where
\begin{eqnarray}
A_{\CP,\T}   = \frac{\Gamma(\Bzb \to \Bz) - \Gamma(\Bz \to \Bzb)}{\Gamma(\Bzb \to \Bz) + \Gamma(\Bz \to \Bzb)}\\
A_{\CP,\CPT} = \frac{\Gamma(\Bz \to \Bz) - \Gamma(\Bzb \to \Bzb)}{\Gamma(\Bz \to \Bz) + \Gamma(\Bzb \to \Bzb)}.
\end{eqnarray}
One can go further than this, as proposed by Bernabeu et al. \cite{Bernabeu:2012ab} and use any two 
different orthonormal basis pairs to classify the decays.  

For example one natural filter basis choice is the determination of the flavour of 
one of the $B$ decays via transitions to a flavour specific final state (e.g. a semi-leptonic decay) 
which filters on $\{B^0, \Bzb\}$.  Another 
choice is that of a neutral meson decaying into a \CP tag final state which filters on $\{B_+, B_-\}$.
Bernabeu et al. proposed the comparison of four sets of processes which satisfy the above
criteria for either \CP, \T, or \CPT operators.  
Here we write these combinations generally in terms of $B$ and $D$ mesons denoted by
$M$, where $\pm$ subscripts refer to the \CP eigenvalue of the \CP filter mode, and the flavour filter
mode is denoted by the decay of particle or anti-particle using the nomenclature of $M^0$ ($\overline{M}^0$).

%
%
Entangled states of neutral meson pairs
produced in \epem interactions at the $\psi(3770)$, \FourS, or \FiveS are given by
\begin{eqnarray}
\Phi &=& \frac{1}{\sqrt{2}} \left( M_1^0\overline{M}_2^0 - \overline{M}_1^0 M_2^0 \right),\\
     &=& \frac{1}{\sqrt{2}} \left( M_{1, +} M_{2, -} - M_{1, -} M_{2, +} \right),
\end{eqnarray}
where $M = \B_{d, s}, \D$,
the subscript $1$, $2$ indicates the first or second meson in the pair,
and the subscript $\pm$ for the \CP filter basis corresponds to the eigenvalue of the \CP eigenstate.
At the time one of the mesons decays the wave function
collapses into a definite state corresponding to either the first ($M^0\overline{M}^0$ or $M_{1, +} M_{2, -}$) or the 
second ($\overline{M}^0 M^0$ or $M_{1, -} M_{2, +}$) ordering.  
The remaining un-decayed meson will propagate through space-time
and mix with its characteristic frequency $\Delta m$ until it too decays.
The possible \CP asymmetries are comparisons of the rates for
\begin{eqnarray}
  \Mzb\to M_- &\text{ vs }& \Mz \to M_-,   \\
  M_+\to \Mz  &\text{ vs }& M_+ \to \Mzb,   \\
  \Mzb\to M_+ &\text{ vs }& \Mz\to M_+,   \\
  M_-\to \Mz  &\text{ vs }& M_-\to \Mzb.   
\end{eqnarray}
The \T asymmetries are comparisons of the rates for
\begin{eqnarray}
  \Mzb\to M_- &\text{ vs }& M_- \to \Mzb,   \\
  M_+\to \Mz  &\text{ vs }& \Mz \to M_+,   \\
  \Mzb\to M_+ &\text{ vs }& M_+\to \Mzb,   \\
  M_-\to \Mz  &\text{ vs }& \Mz\to M_-.   
\end{eqnarray}
The \CPT asymmetries are comparisons of the rates for
\begin{eqnarray}
  \Mzb\to M_- &\text{ vs }& M_- \to \Mz,   \\
  M_+\to \Mz  &\text{ vs }& \Mzb \to M_+,   \\
  \Mz\to M_-  &\text{ vs }& M_-\to \Mzb,   \\
  M_+\to \Mzb &\text{ vs }& \Mz\to M_+.   
\end{eqnarray}
One can compute the corresponding asymmetry observables (the difference over the sum
of the rates) in analogy with Eq.~(\ref{EQ:TASYM}) as a function of the proper 
time difference $\deltat=t_2-t_1$ between the decay of the first meson $t_1$ and that of the 
second meson $t_2$.  This relies on the quantum
coherence of the wave-function of the entangled state over macroscopic scales which, at least
in the case of the \FourS, has been tested experimentally by Belle~\cite{Go:2007ww}.

For example in the \babar measurement the pairing $\Mzb\to M_-$ corresponds to an identified
$\Bz \to \ell^+ X$ (or hadronically tagged) flavour filter decay for the other \B, which
denotes the initial state for the remaining meson as $\Bzb$ at that time.  This
meson subsequently decays via $\B_- \to J/\psi \KS$, which is the \CP-odd filter for 
the decay of a $B_-$. The \T conjugate pairing for this would be the $\B_-$ being tagged on
the opposite side by a $B_+ \to J/\psi \KL$ for the first decay, and for the
second (flavour filter) transition being $\Bzb \to \ell^- X$.

Recently \babar observed \T violation using transitions of entangled pairs of 
neutral \B mesons~\cite{Lees:2012kn}, following the prescription outlined in \cite{Bernabeu:2012ab}.  
The experimental result corresponds to an observation of \T violation in these decays
at the level of $14\sigma$, which is also consistent with the observed level
of \CP violation extracted using the same data-set~\cite{Aubert:2009aw}.  The \belle experiment
should be able to confirm this result, and future flavour factories such as
the upgrade of KEKB and \belle II should be able to improve significantly on the
precision of this test.  The rest of this paper discusses a number of potential \T
violation measurements that can be made at existing experiments, and in the near
future by Belle II.  A few formalities are discussed in terms of the general 
methodology, followed by a discussion of additional \CP filter bases that can be 
used for \T violation measurements.  Following on from this we discuss the 
prospects of performing tests using pairs of neutral \B mesons 
created in the decay of the \FourS, and
the potential to explore \T violation in the charm sector 
using entangled decays of \D mesons at the $\psi(3770)$.
We also mention measurement possibilities for mesons produced in an uncorrelated way.

%
%
As mentioned above, distinct \T violation measurements require the use of two pairs of orthonormal
bases that can be experimentally identified.  \babar and Belle have had great success in 
experimentally separating particle and anti-particle states using flavour tagging algorithms
designed to select flavour specific final states so one can simply follow existing methods
for flavour filtering.  The \CP filter pair basis that has been used until now involves 
neutral \B decays to final states including a charmonium meson ($c\cbar$) 
with a \KS (\CP odd) or a \KL (\CP even) meson.  Here $\{ \KS, \KL\} = \{ -, +\}$ is used to construct an 
approximately orthonormal \CP filter basis, where final states involving a \KL are experimentally
more challenging to identify, resulting in lower signal efficiency (event yield) and lower
purity than the \T conjugate filter with a \KS.

As noted in~\cite{Banuls:1999aj,Banuls:2000ki,Bernabeu:2012ab} and ignoring experimental resolution and dilution 
effects, the decay rate as a function of proper time difference to test \T is given by
\begin{eqnarray}
g_{\alpha, \beta}^{\pm} \propto \left[ 1 + C_{\alpha, \beta}^{\pm}\cos\mathrm\Delta m \deltat + S_{\alpha, \beta}^{\pm} \sin\mathrm\Delta m \deltat\right]\label{eq:timedpendence}
\end{eqnarray}
where $\alpha$ and $\beta$ are the flavour and \CP basis filter decays respectively, the $\pm$ superscript
indicates the time orderings
(i.e. $+$ is the normal ordering of \CP filter decay after flavour filter decay, and $-$ is the inverted order), 
$\mathrm\Delta m$ is the mixing frequency of the meson $M$, \deltat is the proper time difference 
between the two decaying mesons ($\geq 0$ by construction).  The coefficients $S_{\alpha, \beta}^{\pm}$ and 
$C_{\alpha, \beta}^{\pm}$ are given by
\begin{eqnarray}
S_{\alpha, \beta}^{\pm} = \frac{2 Im \lambda}{1 + |\lambda|^2} \text{ and } C_{\alpha, \beta}^{\pm} = \frac{ 1 - |\lambda|^2}{1 + |\lambda|^2},
\end{eqnarray}
where $\lambda = (q/p) \overline{A} / A$.  Here $q$ and $p$ are coefficients parameterising $M^0$-$\overline{M}^0$ mixing
and $\overline{A} / A$ is the ratio of decay amplitudes for the \CP filter.  
Hence $\lambda$ can be used to constrain any weak phase
difference arising from the structure of the \CP filter decay in the overall asymmetry.
Equation~\ref{eq:timedpendence} assumes a lifetime difference $\Delta \Gamma = 0$, which 
can be trivially extended to the more general case.

We note that it is possible to identify other experimentally distinguishable \CP filter bases 
that, unlike $\{ \KS, \KL\}$, are exact.  These arise from the set of $B$ and $D$ decays to final 
states with two spin-one particles, for example pairs of vector (V; $J^P=1^-$) or axial vector (A; $J^P=1^+$) 
particles.  If one performs a transversity analysis of 
such a final state, one can experimentally resolve
\CP even and \CP odd parts as $A_L\equiv A_0$ and $A_{\parallel}$ (even) and $A_\perp$ 
(odd)~\cite{Dunietz:1990cj,Kramer:1991xw}.  Here
$A_L$ is the longitudinal amplitude and $A_{\parallel, \perp}$ are transverse
amplitudes. \footnote{In terms of the helicity basis $A_0\equiv A_L$ is helicity zero $(h=0)$ and the transverse
components are admixtures of $A_{h_\pm}$ with $h_\pm = \pm 1$.  The transverse amplitudes are constructed
as follows: $A_{\parallel}=(A_{h_+}+A_{h_-})/\sqrt{2}$ and 
$A_{\perp}=(A_{h_+}-A_{h_-})/\sqrt{2}$.  It follows that $A_0\equiv A_L$ and $A_{//}$ are \CP-even and $A_\perp$ is \CP-odd.}
These amplitudes can be determined by performing full angular analyses of final states of 
interest. As a result it is important to identify filter decays where there is a significant
transverse component $A_\perp$ (as the \CP odd part of the decay will limit the overall precision on
any measured weak phase).  Thus we study decays in terms of $f(\cos\theta_1, \cos\theta_2, \phi)$
for the helicity basis, or transversity basis angles $f(\cos\theta_{1}, \cos\theta_{tr}, \phi_{tr})$,
where the $\theta_{i}$, (with $i=1$, $2$) are helicity angles given by the decay of 
the positive daughter of the (axial)vector meson with respect to the direction opposite 
to that of the $M (\overline{M})$ in the rest frame of the (axial)vector meson, and $\phi$ is the 
angle between the two decay planes defined by the (axial)vector mesons in the
rest frame of $M (\overline{M})$. The transversity angles $\theta_{tr}$
and $\phi_{tr}$ are the azimuthal and polar angles of one of the decay products 
of the second meson.

Typically the experimental differences between odd and even basis states 
$\{A_\perp,  A_{L \vee \parallel} \}$ of a VV, AV, or AA decay is smaller than that of 
the $\{ \KS, \KL\}$ basis as one reconstructs particles decaying into the 
same final states.  Experimentally one can distinguish between $B_+$ decays
to $A_{L \vee \parallel}$ and $B_-$ decays to $A_\perp$.

The time-dependence of the decaying system can be studied following 
the methodology used for example in $B\to J\psi K^*$ and $B_s\to J\psi \phi$ 
decays~\cite{Itoh:2005ks,Aubert:2007hz,Aaij:2012eq,Aad:2012kba}.
In analogy with Eq. (\ref{eq:timedpendence}),
one can split the time-dependence by time-ordering, flavour, and \CP
filter decay to extract $(S,C)_{\alpha, \beta}^{\pm}$ for each 
amplitude $A_{L, \parallel, \perp}$.

One could also use \CP filter decays to non-\CP eigenstates to perform a \T violation test,
however one would have the added complication of having to distinguish 
between \CP-even and \CP-odd components of said admixture in order 
to identify the orthonormal \CP filter basis.
However, this possibility is not discussed 
further here, but follows the experimental methodology of $B\to \pi^+\pi^-\pi^0$ 
and $B\to a_1\pi$ decays discussed in~\cite{Aubert:2006gb,Kusaka:2007mj,Dalseno:2012hp,Lees:2013nwa}. By using these additional \CP filter 
bases one will be able to perform a number of previously unforeseen
measurements of \T violation in \B and \D systems.

The previous discussion is quite general, the thing that remains is to identify the 
set of pairs of decay channels that can be used to compute $A_\T$, beyond those already
studied.  The measured asymmetry is
\begin{eqnarray}
A_T \simeq \frac{\Delta C^\pm_T}{2} \cos\deltam \deltat + \frac{\Delta S^\pm_T}{2} \sin\deltam \deltat.
\end{eqnarray}
where the parameters $\Delta S^\pm_T$ ($\Delta S^\pm_T$) are linear combinations of
$S^\pm_{\alpha, \beta}$ ($C^\pm_{\alpha, \beta}$) corresponding to the \CP filter sinusoidal 
oscillation amplitudes, and facilitate a set of measurements of angles of unitarity triangles in the 
SM.  
One finds that $|\Delta S^\pm|= 4 Im \lambda / (1+|\lambda|^2)$ and
$|\Delta C^\pm|= 2 (1 - |\lambda|^2) / (1+|\lambda|^2)$ as the deltas are linear combinations
of the previously encountered observables $(S, C)^\pm_{\alpha, \beta}$.  For $b\to c\cbar s$ decays, in the absence of 
direct \CP violation, $|\Delta S^\pm| = 2\sin 2\beta$. Thus we can identify the measured asymmetries with
the underlying weak phase differences in the decay relating to angles of unitarity triangles
in the SM.
For a non-zero asymmetry to be manifest we require a weak phase difference between the two rates
used to construct the asymmetry.
In the case of kaons this phase difference is introduced via the 
mixing process.  For the decays discussed in this paper a phase difference could originate from
any (or all) of the following: flavour filter; \CP filter; the mixing transition.  Given
our knowledge of \CP violation in \B decays we expect that the phase difference is manifest
in the interference between mixing and decay amplitudes related to the \CP filter. 
\footnote{We note that hadronically tagged events can manifest 
a small level of \CP violation from the interference between Cabibbo allowed and Cabibbo
suppressed amplitudes, however this is a second order effect~\cite{Long:2003wq}.}

%
%
For $\Bz(\Bzb)$ mesons, decays to flavour specific final states such as $X\ell^\pm \nu$
constitute the flavour tag decay part of the problem.  The other decay is into a \CP odd or even 
eigenstate which has an experimentally distinguishable conjugate.  

The measurements performed by \babar used a
set of charmonium plus \KS decays ($c\cbar\KS$) and their \CP-conjugated partners
($c\cbar\KL$) as the \CP filters which are paired with flavour tag filters described above.  
These $c\cbar\Kz$ states are theoretically clean to interpret within the SM hence the importance of 
studying them. 
It is also possible to perform such a measurement with \emph{any} pair 
of final states that exhibit time-dependent \CP violation such as the 
\CP conjugate pairs $\B \to (\eta^\prime, \phi, \omega) \KS$ and $(\eta^\prime, \phi, \omega) \KL$.
These can be used to provide alternative measurements of $\beta$ to complement the existing tree
level determination obtained by \babar.
\CP tag filters such as the $b\to s$ loop transitions $\eta^\prime \KS$ and $\eta^\prime \KL$ are ideal choices for
another \T violation measurement as
physics beyond the SM may affect the 
asymmetries via non-SM loop amplitudes contributing to the final state.  
It is important to test
the level of \T violation measured using tree process \CP filters against that found using loops as a priori the
nature of new physics is unknown and may be manifest here.
It is also possible to study \T symmetry invariance using $b\to u$ (tree dominated) \CP filter transitions such as
$\B \to \rho\rho$ and its axial vector counterparts.  Here the decay to $\rhoz\rhoz$ is of
most interest as there are indications of a sizeable \CP odd component in that final state (the $B_-$ filter). 
The weak phase difference is manifest in the interference between mixing and decay 
amplitudes of the \CP filter.
Similarly one can test
$b\to d$ (loop) transitions via $B\to D^{*+}D^{*-}$ decays, which also measures $\beta$. 
The loop dominated $b\to s$ decay
$\Bz \to \phi K^*$ is of particular interest as a \CP filter because there is a sizeable \CP-odd component ($B_-$ filter)
to compare with the dominant \CP-even ($B_+$ filter) part of the decay.
This set of experimental tests can be used to independently compare
\T, \CP, and \CPT symmetry violation/invariance for $b\to u,\,c,\,d$, and $s$ transitions in order to verify if the 
SM holds up to scrutiny.  The time-dependent \CP asymmetry parameters have already been
determined for these modes, so in a sense half of the job has been done by 
the \BFs.  These experiments have successfully confirmed the real benefit of 
studying \B decays is the miracle that \CP violation is manifestly large for these systems.
For $b\to c,\,d$, and $s$ quark transitions $\Delta S^\pm$ are 
measures of $\mp 2 \sin2\beta$, and they are related to the unitarity triangle angle
$\alpha$ for $b\to u$ transitions.
Using the available data we estimate (see Table~\ref{tbl:estimates})
the precision $\sigma(\Delta S^\pm)$ that the current \BFs and Belle II can be 
expected to reach in some of the aforementioned modes.
Belle II should be able to observe \T violation at the SM rate in these \CP filter basis channels.
\begin{table}[!ht]
\caption{Estimated sensitivities on $\sigma(\Delta S^\pm)$ for the \BFs and Belle II.}\label{tbl:estimates}
\renewcommand{\arraystretch}{1.2}
\begin{tabular}{c|cc}\hline\hline
\CP Filter basis pair & \BFs & Belle II \\ \hline
$\eta^\prime K^0_{S/L}$ & 0.6 & 0.08 \\
$\phi K^*$              & 1.1 & 0.13 \\
$\eta K^0_{S/L}$        & 1.8 & 0.17 \\
$\omega K^0_{S/L}$      & 2.0 & 0.22 \\
$D^* D^*$               & 2.0 & 0.29 \\ \hline\hline
\end{tabular}
\end{table}

It may be possible to extract information from $\B_s$ meson pairs collected at the \FiveS,
however it should be noted that current vertex detector technology is insufficient to resolve
individual neutral meson oscillations in this system at an \epem collider experiment 
such as Belle II, so one must rely on the $\Delta \Gamma$ modulation of the neutral meson 
oscillation to obtain information related to $A_T$.  $\B_s$ mesons are produced by 
hadronic collisions at the LHC, hence it would not be possible to perform a \T-violation 
measurement using entangled pairs at the LHC experiments.  However, one could perform a Kabir 
asymmetry test by studying flavour tagged $\B_{s}$ decays, which would allow the 
measurement of $A_{\CP, \T}$ ($A_{\CP, \CPT}$).   These are measurements of the 
difference in probabilities between $\Bsb \to \Bs$ ($\Bs \to \Bs$) and that 
of $\Bs \to \Bsb$ ($\Bsb \to \Bsb$), which could be done as a function of proper time using 
currently available techniques.  ATLAS, LHCb and the Tevatron experiments have demonstrated their capability to perform
such a measurement through their studies of time-dependent \CP asymmetries in $\B_s\to J/\psi \phi$~\cite{hfag}.
Hence one can use the same flavour tagging techniques as the initial flavour filter, 
and use semi-leptonic \Bs decays as the second flavour filter.

%
%
Now we can turn to the issue of charm decays where we recently urged the experimental community
to embark upon the systematic study of time-dependent \CP asymmetries to parallel the work
of the \BFs since 1999~\cite{Bevan:2011up}.  This was in part motivated by the opportunity to make such 
measurements given the availability of data from LHCb and the promise of more to come
from future flavour factories. Following our previous paper there have been 
tantalising indications of a non-zero direct \CP asymmetry in $D\to K^+K^-$ and $D\to \pi^+\pi^-$ decays
from LHCb and CDF~\cite{Aaij:2011in,Aaltonen:2012qw}, however current data are consistent with 
\CP conservation~\cite{hfag}.
The underlying physics
regarding the production of neutral \D mesons in \epem collisions at the $\psi(3770)$ is a 
direct parallel of the production of \B mesons at the \FourS.  We note that as a consequence
of this, one can re-use the measurement technique of Ref.~\cite{Bernabeu:2012ab} adopted by \babar and apply this to \T violation
searches in charm mesons,
with the caveat that the lifetime difference matters for charm as $y = \Delta \Gamma/2\Gamma$ is 
non-zero.  In analogy with our observations for \B decays
we note that one could test \T at an asymmetric \D factory running 
at the $\psi(3770)$ using a number of different final states.
For example one can study \T invariance for semi-leptonic and hadronic flavour filter and $c\to d$ 
and $c\to s$ \CP filter transitions at leading order.
While it is, in principle, possible to access $c\to u$ real and $c\to b$ virtual transitions 
from the second order, CKM suppressed, loop contributions, any results would be difficult to 
interpret in terms of the CKM matrix as one first needs to constrain the phase of charm mixing.

In the SM, the level of \CP violation in charm is expected to be small so, unless \T violation
arises from other mechanisms, any experimental test of \T
symmetry invariance would require a high degree of systematic control
(and careful design to minimise systematic uncertainties), and a large integrated luminosity
of data collected via $\epem \to \psi(3770) \to \DzDzb$.  Thus one expects an asymmetry
compatible with zero unless there are significant new physics contributions. 
Given the sample sizes of experiments
under discussion, with a few \invab, one would not expect to be sensitive to the 
level of \T violation compatible with the SM --- however that does mean that if one
were to perform a \T symmetry measurement and observe \T violation, that would have
to result from physics beyond the SM.  
Experimentally, the dilution effect present in
determining flavour tag $B$ states is absent for semi-leptonic flavour tagged neutral $D$ mesons 
as there are only primary leptons
produced in $\Dz(\Dzb)\to X_s \ell \nu$ decays, and as both \D mesons are reconstructed 
with a well known initial set of conditions, there should be very little background
present (at least in final states without a \KL meson) with which one could study 
\T symmetry invariance.  The other experimental issue of relevance is that of detector 
resolution.  Using vertex detector technology accessible today, with a center of mass
boost factor $\beta\gamma$ relative to the laboratory frame of at least about 0.3-0.4, 
it should be possible for a $\tau$-charm factory to study \T symmetry invariance in a 
wide range of final states including $h^+h^-\KSL$. \CP tag decays such as 
$\Dz \to \KSL \piz$ will be difficult to reconstruct, but
should be experimentally accessible in terms of \CP violation measurements, however the lack of charged particles 
originating from the $D$ decay vertex means that the tag mode with $\CP=+1$ would 
have to be reconstructed via \piz Dalitz decays or photon conversion processes.  However
equivalent measurements with $\Dz \to \KSL (\omega,\, \eta,\, \eta^\prime, \, \rho^0,\, \phi,\, f_0,\, a_0)$
are also possible, where these are expected to be less experimentally challenging than the $\KSL \piz$
combination.  
Studies of $\Dz \to \KS\KS\KL$ and the \T-conjugate $\KL\KL\KS$
mode could be used to explore the behaviour of this symmetry for $W$ exchange amplitudes, 
assuming that one can reconstruct the decay vertex of the latter mode and isolate a clean
signal with reasonable efficiency.  
These allow one to probe \T symmetry invariance using $c\to d$ and $c\to s$ \CP filters
to complement the set of measurements in $b$ quark transitions.
A clean theoretical interpretation of such a measurement assumes negligible long distance 
 contamination (i.e. strong force induced transitions) to the underlying weak structure 
of interest.
In analogy with the \Bs scenario, the \BFs and LHC experiments can measure $A_{\CP, \T}$ 
($A_{\CP, \CPT}$) as a function of the proper time difference.   These are measurements 
of the difference in probabilities between $\Dzb \to \Dz$ ($\Dz \to \Dz$) and that
of $\Dz \to \Dzb$ ($\Dzb \to \Dzb$).

%
%
In summary, the recent observation of \T violation using flavour filters with $b\to c$ \CP filter 
transitions by \babar raises an interesting (and old) question: what discrete symmetries are respected 
by the weak force? This paper introduces new \CP filter bases that can be used to perform \T symmetry invariance
tests and outlines a number of measurements that can be made to test \T.  These cover all
kinematically accessible \CP filter quark transitions, i.e. for \B decays one has $b\to u$, $d$, $s$ and $c$ 
transitions (to complement the existing $b\to c$ tests), and for \D decays one can probe
$c\to d$ and $c\to s$ transitions.  While $c\to u$ transitions can be tested in principle,
these are unlikely to ever be measurable.
One expects sizeable \T violation in the various \B decay final states to balance
the open form of the $bd$ Unitarity triangle.  These are alternative measurements
of the unitarity triangle angles $\alpha$ and $\beta$.  Small effects (essentially zero
within experimental precision that would be achievable with facilities under 
consideration today) are expected for \D decays.  
The $b\to d$ and $s$ transition effects should be comparable in magnitude to those 
reported by \babar for $b\to c$ transitions as they are a measure of the unitarity triangle
angle $\beta$, whereas one expects a smaller value for $b\to u$ transitions which are a measure of
$\alpha$ in the SM.
By measuring this set of decays one would be able to constrain leading and 
higher order \T violation contributions in the SM to complement 
the set of \CP violation constraints reported by the \BFs since 1999. 
To do this one requires high statistics data samples from future \epem based (asymmetric energy) 
\B and $\tau$-charm flavour factories operating at the \FourS and $\psi(3770)$,
respectively.  The data samples recorded by \babar and Belle
provide a starting point for detailed exploration of \T violation in \B decays before
Belle II starts taking data later this decade.   Estimates of achievable sensitivities 
by those experiments are given.
The full set of measurements indicated in this article would probe \T violation using 
tree and loop \CP filter transitions for both up and down type quarks. 
It is also worth remembering that one can probe (and over-constrain) the
Kobayashi-Maskawa mechanism and the CKM matrix in terms of the 
set of discrete symmetries \T, \CP, and \CPT.
Hence one can systematically probe the behaviour of the weak force in terms
of these symmetries to complement almost five decades of study with regard to \CP.
The LHC experiments can measure $A_{\CP, \T}$ and $A_{\CP, \CPT}$ in \Bs and \Dz 
systems.  These observables can also be measured by the \BFs for the $\Dz$ system.

\section{Acknowledgments}
We wish to thank Jose Bernabeu, Fernando Martinez-Vidal and Brian Meadows
for useful discussions during the preparation of this paper.

\end{document}